\documentclass[12pt]{article}
\usepackage{etex}
\usepackage{epsfig}
\usepackage{epsf}
\usepackage{latexsym}
\usepackage{amsfonts,amssymb,amsmath} 
\usepackage{mathtools}
\usepackage{color}
\usepackage[a4paper,vmargin=3cm,hmargin=2.1cm]{geometry}
\usepackage{tikz}
\epsfverbosetrue
\newcommand{\de}{\hbox{\rm{d}}}

\newcommand{\dpp}{\vcentcolon}
\newcommand{\bb}{\begin{eqnarray}}
\newcommand{\ee}{\end{eqnarray}}
\newcommand{\eee}{\nonumber\end{eqnarray}}
\newcommand{\qq}{\quad}

\begin{document}

\thispagestyle{empty}

\begin{center}
${}$
\vspace{3cm}

{\Large\textbf{The Hubble diagram in a Bianchi I universe}} \\

\vspace{2cm}

{\large

Thomas Sch\"ucker\footnote{
CPT, Aix-Marseille University, Universit\'e de Toulon, CNRS UMR 7332, 13288 Marseille, France
\\\indent\qq
thomas.schucker@gmail.com }
}

\vspace{2cm}

{\large\textbf{Abstract}}
\end{center}
The Bianchi I metric describes a homogeneous, but anisotropic universe and is commonly used to fit cosmological data. A fit to the angular distribution of 740 supernovae of type Ia with measured redshift and apparent luminosity is presented. It contains an intriguing, yet non-significant signal of a preferred direction in the sky. The Large Synoptic Survey Telescope being built in Chile should  measure some 500 000 supernovae within the next 20 years and verify or falsify this signal.
\vspace{2cm}

\vspace{2cm}

\noindent
Contribution to the 2014 Maui  conference on Subfactor
Theory in Mathematics and Physics,
dedicated to Vaughan Jones' 60th birthday, \\
to appear in the  "Proceedings of the Centre for Mathematics and its
Applications",\\
 editors:
David Penneys and Scott Morrison

\vspace{2cm}
\vskip 2truecm

\section{Introduction}

Vaughan and I arrived in Geneva in 1974. Vaughan started a thesis on the foundations of quantum mechanics at the Ecole de Physique where I continued my under-graduate studies. Since then physics has remained one of our common pleasures. Here is an anecdote from 1977. We were sitting together in Jean-Pierre Eckmann's lectures on quantum field theory. I had just followed a seminar at CERN announcing the discovery of the the fifth quark, the `bottom' quark, and full of enthusiasm I told Vaughan about it during a lecture break. His reaction was: `So they added another epicycle.' Today we have six quarks and they are among the basic building blocks of the standard model of particle physics.

My birthday present for Vaughan is a popular epicycle added to today's standard model of cosmology. This epicycle weakens the cosmological principle, which postulates the 3-space of our universe to be maximally symmetric.

Let me illustrate the cosmological principle by a story that is set in the Black Forest, on a farm in Oberwolfach, home to Traudi, her owner's preferred cow. Traudi is ill and the veterinarian is helpless. In his despair, the farmer spares no effort and calls on a physician, more expensive and as helpless as his colleague. The farmer has a nephew with a PhD in biology and asks him to see Traudi, again without success. Finally, seeing a theoretical physicist on his way to a conference, he asks him for help. The physicist sits down next to Traudi, pulls out his note pad and starts calculating. During hours the farmer watches the physicist's intense concentration from a respectful distance and feels a timid ray of optimism. He pulls closer, caresses Traudi between the horns and asks: `is there hope?'. `Indeed there is', replies the physicist, `I just solved the case of the spherical cow.'.

Let me motivate the weakening of the cosmological principle by an analogy comparing cosmology and the description of the earth's surface. In good approximation, this surface is maximally symmetric, i.e. a 2-sphere. We observe a breaking of this symmetry of the order of one in a thousand by the geography, for example by Mount Everest, $8.8\, {\rm km}\cdot 2\pi /(40\,000\,{\rm km})\,\approx\, 1.4\cdot10^{-3}.$ Of course we do not try to describe these geographic deviations by a simple geometric model. But there is a second breaking of the maximal symmetry, of the order of 3 in a thousand, that we call geometric. Indeed this breaking admits a simple geometric description by an oblate ellipsoid. Our polar radius is about 21.3 km shorter than the equatorial ones, $21.3\, {\rm km}\cdot 2\pi /  (40\,000\,{\rm km})\,\approx\, 3.3\cdot10^{-3}.$

\begin{figure}[h]
\begin{center}
\includegraphics[width=9.5cm, height=6.1cm]{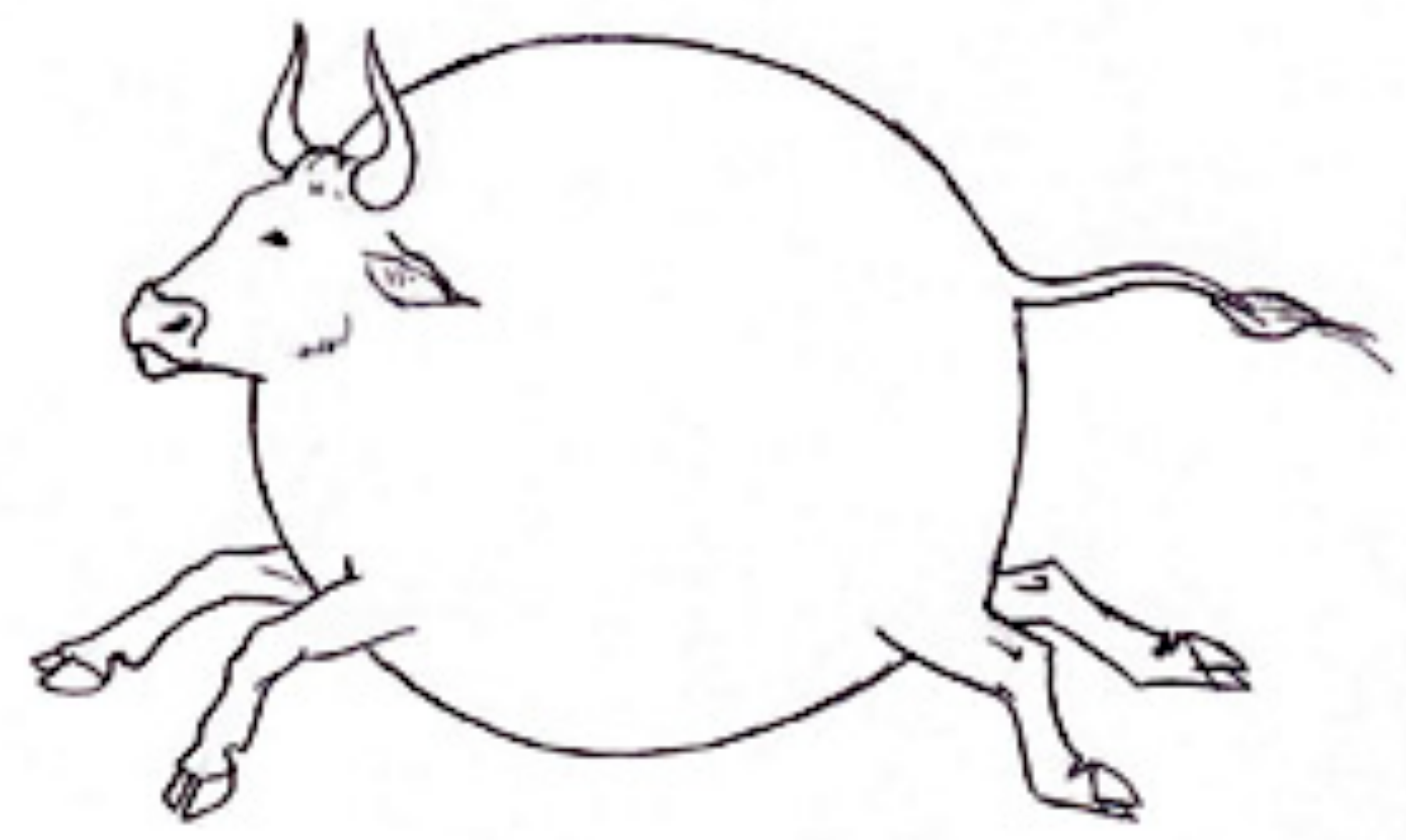}
\caption[]{A spherical cow}
\label{fig1}
\end{center}
\end{figure}

The Robertson-Walker metric of the cosmological standard model has maximal spatial symmetry. The Bianchi I metric, that we consider in the following for its calculational simplicity, is obtained from the flat Robertson-Walker metric by giving up isotropy. It is true that we observe anisotropies of the order of $10^{-5}$ in the cosmic micro-wave background and even stronger anisotropies in the form of galaxy clusters and big voids. But we take these to be geographic deviations and do not try to model them by a simple geometry. Our motivation for using the Bianchi I metric is that it might describe a new breaking of maximal symmetry, of geometric type, in the distributions of supernovae.

\section{Isometry groups}

The isometry group of any (pseudo-) Riemannian manifold of dimension $n$ is a Lie group of dimension less than or equal to $n\,(n+1)/2$. There are three Riemannian manifolds of maximal symmetry, the sphere, Euclidean space and the pseudo-sphere. Likewise there are three spacetimes of maximal symmetry, de Sitter space, Minkowski space and anti de Sitter space. However they are too rigid to be of use in modelling cosmology. Therefore the cosmological principal only postulates maximal symmetry of 3-dimensional subspaces of simultaneity in spacetime. When these are Eucidean spaces we have the flat Robertson-Walker spacetime with metric 
\bb \de \tau^2= \de t^2-a^2(t)\,(\de x^2+\de y^2+\de z^2),\ee
where $a(t)$ is a positive function, called scalefactor. In two instead of three space-dimensions, you can visualize this Robertson-Walker spacetime as an infinite rubber sheet that is stretched in all directions by the same time-dependent factor $a(t)$. The 4-dimensional flat Robertson-Walker spacetime has six isometries, three rotations and three translations. 

Our aim is to break one or more of these isometries. Breaking only one isometry is impossible:

{\bf Theorem} (Fubini 1903) The isometry group of a Riemannian manifold of dimension $n\ge 3$ cannot be of dimension $n\,(n+1)/2\,-\,1$.

Therefore we are looking for 3-spaces with four or three isometries. The second case has been classified by Bianchi. There are nine such spaces, Bianchi I to IX. The first one is the simplest one, leading to a spacetime metric with three scalefactors:
\bb \de \tau^2= \de t^2-a^2(t)\,\de x^2-b^2(t)\,\de y^2-c^2(t)\,\de z^2.\ee
It is invariant under three translations. To simplify we take $a(t)\equiv b(t)$, thereby adding a fourth isometry, the rotations around the $z$-axis.

 \section{The observed Hubble diagram}
 
 There are good reasons to believe that all supernovae of type Ia are standard candles, i.e. they all radiate light with the same absolute luminosity (at peak emission). This value is $L=(1\pm0.1)\cdot 10^{35}$ Watt. Today we have a catalogue of 740 such supernovae, the `Joint Light curve Analysis' (JLA) \cite{jla}. For each supernova the redshift $z$, the apparent luminosity $\ell$ and the direction in the sky with respect to the equatorial plane of the Earth, i.e. declination $\theta $ and right ascension $\varphi$, are measured. 
 
 The redshift is obtained from the Doppler effect on atomic spectra caused by the recession of the galaxy hosting the supernova with respect to our galaxy. It is by definition $z\dpp=(T_D - T)/T$ where $T$ is the period of any spectral ray of any atom measured on earth and $T_D$ the same Doppler shifted period in the light coming from the supernova. All $z$s in the JLA catalogue are positive, indicating a recession of the host galaxies. (Negative $z$s would mean blueshifts indicating a contracting universe.) The largest redshift in the catalogue is $z=1.3$.
 
 The apparent luminosity is the power of the light collected per unit surface of the detector. It is given by $\ell = L\,\de \Omega _{-1}/\de S_0\,/(z+1)^2$, where $\de\Omega _{-1}$ is the infinitesimal solid angle of the light beam at emission and $\de S_0$ the corresponding infinitesimal surface of the beam perpendicular to the  beam at detection. The two negative powers of $z+1$ take into account the energy loss of every photon caused by the redshift and the loss of power caused by the change in the time unit between emission time $t_{-1}$  and detection time $t_0$ today.
 
 The Hubble diagram is the plot of the catalogue data in the $z$-$\ell$ plane. Indeed under the assumption of the cosmological principle, 
the redshift and apparent luminosity of a standard candle is independent of its direction in the sky. 

\section{Computing the Hubble diagram}

Suppose we know the scalefactors $a(t)$ and $c(t)$ in the Bianchi I metric. Suppose that we neglect peculiar velocities of our and all host galaxies. Then their (`comoving') trajectories in spacetime are given by the time-like geodesics $t=\tau$, our proper time, $x,\,y$ and $z$ constant. Likewise, the trajectories of the photons are light-like geodesics. They are straight lines in the isotropic case, not so in the Bianchi I metric (unless their initial direction is a principal direction, the $z$ axis or any direction perpendicular to it). Therefore their generic computation, given in \cite{stv}, is not reproduced here. Let us note that the time derivative of the two scalefactors cannot differ much because otherwise distant objects would show a drift, that is their generic position in the sky would change with time. Such a drift is unobserved today and the Hubble stretch $\epsilon(t)$ defined by $\de a/\de t=\dpp\de c/\de t\,(1+\epsilon)$ must be small. 

From the knowledge of the light-like geodesics, one then computes the redshift and the apparent luminosity as a function of the time of flight of the photon $t_0-t_{-1}$ and its incoming direction. Unfortunately the incoming photons cannot tell us their time of flight. It has to be eliminated leading to the parametric plot in the $z$-$\ell$ plane for any given direction. 

\section{Computing the scalefactors}

The two scalefactors $a(t)$ and $c(t)$ are computed by integrating the Einstein equation with a positive cosmological constant $\Lambda $ and pressure-less i.e. collision-less matter, the comoving galaxies, intergalactic matter and dark matter, with a total mass density $\rho$. The translation invariance of the metric implies also that the mass density can only depend on time. To get a unique solution of the Einstein equation, it is sufficient to know the `final' values today $\rho (t_0) =\dpp \rho _0$ together with the Hubble constant $H_0\dpp = \de a(t_0)/\de t/a(t_0)$ and the Hubble stretch today $\epsilon_0 \dpp=\epsilon (t_0)$. Again let me refer to \cite{stv} for the details of the integration. 

\section{The fit }

\begin{figure}[h]
\begin{center}
\includegraphics[width=15cm]{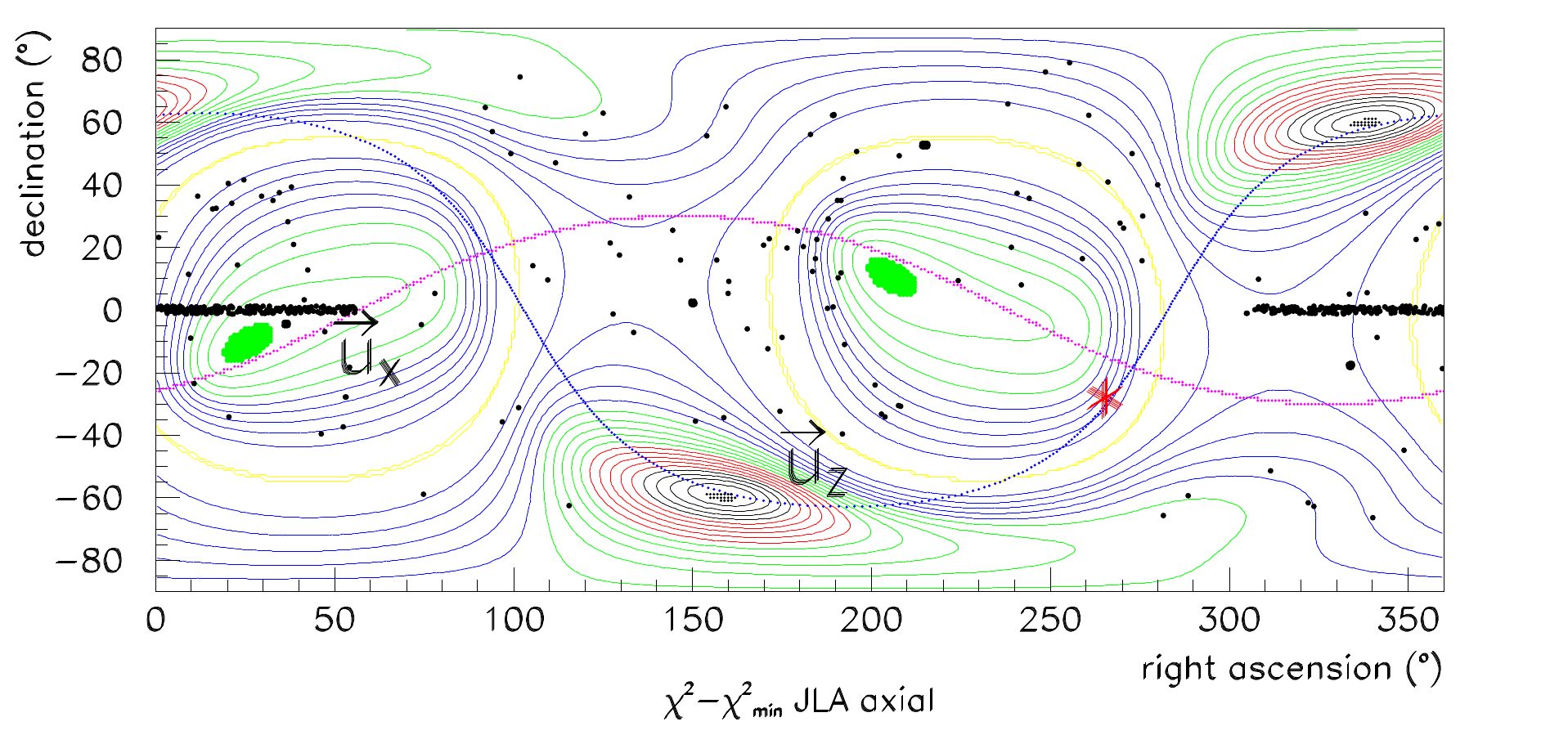}
\caption[]{Confidence level contours of privileged directions in arbitrary color
  codes for Bianchi I spacetimes with axial symmetry. Black points represent supernova positions. Note the accumulation  of supernovae in the equatorial plane of the Earth. The blue line is the
  galactic plane and the purple line is the plane transverse  to the main privileged
  direction (gray speck). The red star is the direction towards our  galactic center.}
\label{fig2}
\end{center}
\end{figure}

Now we are ready to fit the computed Hubble diagram to the JLA data by adjusting the following parameters: $\rho _0,\, H_0,\, \epsilon_0$ and the privileged direction (the $z$ axis) $\theta _p,\, \varphi _p$. Some 100\,000 CPU hours later we  find a weak signal \cite{stv} at $1\,\sigma $ confidence level: 
\bb \theta _p= -60 ^\circ\pm 11^\circ,\qq
\varphi _p=158^\circ\pm29^\circ,\qq
\epsilon_0=-1.7\,\%\pm 1.3\,\%.\ee
Figure \ref{fig2} gives more details. In particular the privileged direction is in our galactic plane and almost perpendicular to the direction towards our galactic center.

This signal may well be a statistical fluctuation. 

In 2014 the construction of the Large Synoptic Survey Telescope began in Chile. It should start taking data in 2021 at a breathtaking rate of $50\,000$ supernovae of type 1a per year and operate during 10 years. We expect that after one year of operation the above three error bars reduce to $ \pm12^\circ,\,\pm 7^\circ,\,0.06\,\%$, after 10 years to $ \pm4^\circ,\,\pm 2^\circ,\,0.02\,\%$. Therefore we should know in some 10 years if our universe prefers a rugby ball to a spherical cow.

I would like to take the opportunity to thank Sarah, Arnaud, Dave and Scott for their  friendly and efficient organisation of the Maui conference.\\
\begin{center}
\vspace{.1cm}
Bon vent \`a Vaughan\\
\vspace{,1cm}
\end{center}


\begin{thebibliography}{10}
  \bibitem{jla}  
 M.~Betoule {\it et al.} [SDSS Collaboration],
  ``Improved cosmological constraints from a joint analysis of the SDSS-II and SNLS supernova samples,''
  Astron.\ Astrophys.\  {\bf 568} (2014) A22
  [arXiv:1401.4064 [astro-ph.CO]].
  \bibitem{stv}
  T.~Sch\"ucker, A.~Tilquin and G.~Valent,
  ``Bianchi I meets the Hubble diagram,''
  Mon.\ Not.\ Roy.\ Astron.\ Soc.\  {\bf 444} (2014) 3,  2820
  [arXiv:1405.6523 [astro-ph.CO]].
  
 
\end{thebibliography}
\end{document}